\documentclass[final,5p,times,twocolumn,authoryear]{elsarticle}
\usepackage{natbib}
\usepackage{lineno}
\usepackage{graphicx,color}
\usepackage{float}
\usepackage{ulem}

\usepackage{amssymb}
\usepackage{amsmath}

\journal{}

\begin{document}

\begin{frontmatter}

\title{Time-resolved sedimentation of dense potato-starch suspensions measured by optical coherence tomography} 

\author[1]{Tasuku Saiki}
  \author[1]{Hiroaki Katsuragi}

\affiliation[1]{organization={Department of Earth and Space Science, 
  The University of Osaka},
  addressline={1-1 Machikaneyama-cho}, 
  city={Toyonaka},
  postcode={560-0043}, 
  country={Japan}}

\begin{abstract}
We demonstrate optical coherence tomography (OCT) as a measurement technique for dense, optically opaque suspensions. Conventional optical methods cannot access the interior of such suspensions. OCT resolves individual potato-starch particles (${\sim}20~\mathrm{\mu m}$) as distinct scatterers, even though the suspension appears opaque to the eye. By tracking the vertical centroid position of the particle-laden layer $\langle Z \rangle(t)$ and the supernatant boundary $Z_\mathrm{sup}(t)$ in the same OCT image sequence, we obtain the instantaneous settling velocity $V(t)$ and the time-evolving effective volume fraction $\phi_\mathrm{eff}(t)$ simultaneously and continuously in time. To our knowledge, this is the first measurement to combine settling velocity and particle concentration into a single continuous trajectory within one sedimentation run. Conventional batch measurements yield only one velocity value per run. We applied this method to dense potato-starch suspensions, varying the initial volume fraction $\phi_0$ from 0.30 to 0.50 and the solvent density $\rho_\mathrm{L}$ from 1.0 to $1.3{\times}10^3~\mathrm{kg~m^{-3}}$ using aqueous sodium polytungstate solutions. The normalized velocity $V/V_\mathrm{Stokes}$ plotted against $\phi_\mathrm{eff}$ collapses onto a common trend consistent with both the Krieger--Dougherty model and the Richardson--Zaki law over $\phi_\mathrm{eff} \simeq 0.30$--$0.52$, confirming that the method captures physically reasonable hindered-settling behavior. These results establish OCT as a viable tool for probing internal dynamics in dense suspensions that were previously inaccessible to optical measurement.
\end{abstract}

\end{frontmatter}

\section{Introduction}
\label{sec:intro}

Dense suspensions are often optically opaque. This makes it difficult to observe particle motion inside the suspension during sedimentation. Classical theory describes hindered settling through models such as the Kynch kinematic theory~\citep{Kynch:1952}, the Batchelor correction~\citep{Batchelor:1972}, the Krieger--Dougherty (K-D) model~\citep{Krieger:1959}, and the Richardson--Zaki (R-Z) law~\citep[originally published in 1954]{Richardson:1997}. These models are well established. But the underlying data usually come from external observation of the sedimentation front, not from direct internal measurement.

Time-resolved measurements inside dense suspensions ($\phi \gtrsim 0.3$) are difficult to obtain~\citep{Davis:1985,Stickel:2005,AlNaafa:1992}. Conventional batch experiments track only the descending front from outside the container. This gives a single run-averaged velocity, not a continuous measurement. X-ray phase-contrast tomography can resolve individual particle trajectories~\citep{Ruhlandt:2019}. However, it requires sufficient X-ray contrast between particle and fluid. This contrast is small for organic particles in water. The method is therefore limited to a narrow range of systems and volume fractions. A measurement technique that works directly inside optically opaque, aqueous suspensions is still needed.

This need is not merely technical. Some dense suspensions show anomalous rheology under fast deformation. Dense potato-starch suspensions exhibit discontinuous shear thickening (DST)~\citep{Brown:2014} and impact-activated solidification~\citep{Waitukaitis:2012,Egawa:2019}. These phenomena appear under rapid, impact-driven deformation. Sedimentation is a slow, gravity-driven process instead. The shear rate during sedimentation is expected to stay well below the onset shear rate for DST. So the anomalous rheology likely does not affect quasi-static sedimentation. This expectation has not been checked by direct measurement. Dense, non-Brownian sedimentation is also known to show large hydrodynamic velocity fluctuations~\citep{Nicolai:1995,Guazzelli:2011}. A direct, time-resolved measurement can settle both points rather than leave them as assumptions.

Optical coherence tomography (OCT) can meet this need. OCT was originally developed for cross-sectional imaging of biological tissue. It uses low-coherence interferometry~\citep{Huang:1991}. It achieves micrometer-scale depth resolution. It is contact-free and non-invasive~\citep{Fercher:2003}. OCT can also image through turbid media. This property makes it suitable for observing particle dynamics inside suspensions that appear opaque to the eye. OCT has recently been applied in soft matter research~\citep{Amini:2025}. It has also been applied in fluid mechanics, including flow velocimetry~\citep{Buchsbaum:2015,Mujat:2013,Zhou:2016}. However, its use for tracking sedimentation inside dense, particle-laden suspensions has not been demonstrated.

Here, we apply OCT to measure sedimentation in dense potato-starch suspensions at initial volume fractions $\phi_0 = 0.30$--$0.50$. We track two quantities from the same OCT image sequence. One is the centroid of the particle layer. The other is the boundary of the growing supernatant layer. This gives the settling velocity $V(t)$ and the effective volume fraction $\phi_\mathrm{eff}(t)$ simultaneously and continuously in time. To our knowledge, this is the first measurement to combine these two quantities into a single continuous trajectory within one sedimentation run. We compare the normalized velocity $V/V_\mathrm{Stokes}$ against $\phi_\mathrm{eff}$ with the K-D model~\citep{Krieger:1959} and the R-Z law~\citep{Richardson:1997}. This comparison tests two things. First, whether OCT-based measurement gives physically reasonable hindered-settling behavior. Second, whether any signature of the anomalous rheology appears in this quasi-static regime.

\section{Experiments}
\label{sec:exp}

\subsection{OCT imaging}
\label{sec:exp:OCT}

A microscope-mounted OCT system (LUMEDICA OQ Pathscope mounted on Amscope microscope with a 4× objective lens) was used for all measurements. The system operates at a wavelength of $850~\mathrm{nm}$ and with a pixel size of approximately $1.9~\mathrm{\mu m/pixel}$. Images were acquired at $10~\mathrm{fps}$ with a frame size of $512{\times}512$ pixels. Namely, the spatial and temporal resolutions are therefore set by the pixel size ($1.9~\mathrm{\mu m}$) and the frame interval ($0.1$~s), respectively. Both resolutions are well below the relevant length and timescales of particle sedimentation in this study. Each sedimentation typically lasted several hundred seconds, although the duration depended on the experimental conditions.

\begin{figure*}
  \centering
  \includegraphics[width=\linewidth]{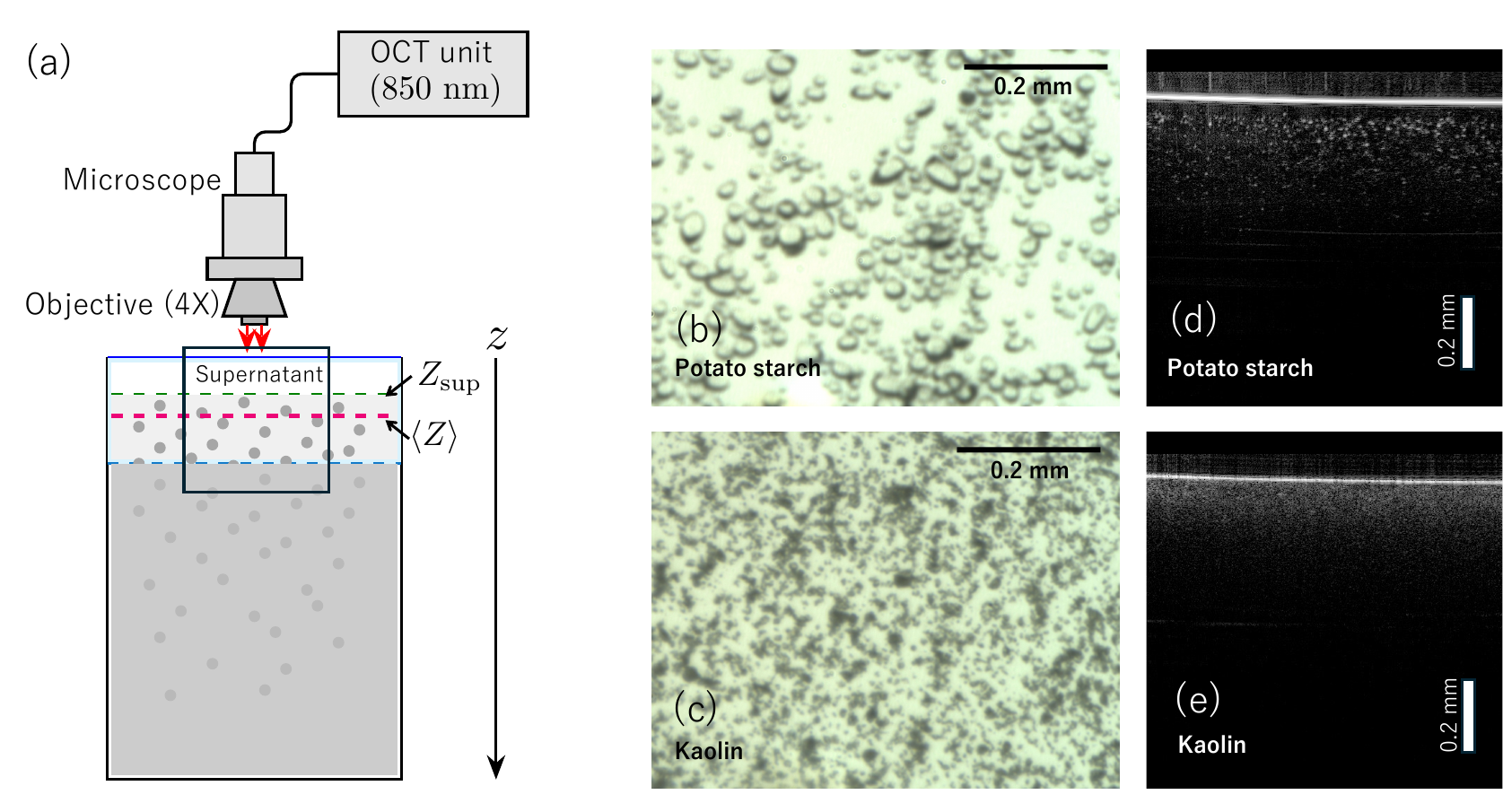}
  \caption{(a)~Schematic of the experimental setup. (b)~Direct microscope image (not by OCT) of potato-starch particles. (c)~Direct microscope image (not by OCT) of kaolin particles. (d)~OCT image of a dense potato-starch suspension (vertical cross-section). (e)~OCT image of a kaolin suspension (vertical cross-section), showing a speckle pattern rather than individually resolved particles.}
  \label{fig:setup}
\end{figure*}

The OCT probe was directed downward (Fig.~\ref{fig:setup}(a)), imaging a vertical cross-section of the suspension just below the liquid surface. The observable depth is approximately $1~\mathrm{mm}$ from the surface. We define the $z$-axis as the vertical direction pointing downward, and the $x$-axis as the horizontal direction within the imaged cross-section. Since only relative displacement (i.e., velocity) is used in the analysis, no specific origin is fixed. The OCT image is therefore a two-dimensional $x$--$z$ slice of the three-dimensional suspension, analogous to a laser-sheet cross-section in particle image velocimetry (PIV).

\subsection{Materials and suspension preparation}
\label{sec:exp:materials}

Potato starch (katakuriko in Japanese) was used as the particle phase. The particles have a broad size distribution. Manual measurement of 10 randomly sampled particles from the microscope image (Fig.~\ref{fig:setup}(b)) using ImageJ gives a mean diameter of $24 \pm 10~\mathrm{\mu m}$ (mean $\pm$ standard deviation), consistent with the nominal value $D_\mathrm{p} \simeq 20~\mathrm{\mu m}$. Given the small sample size and the resulting large scatter in this manual measurement, we retain the nominal value $D_\mathrm{p} \simeq 20~\mathrm{\mu m}$ for the calculation rather than the measured mean. The density of the particles is $\rho_\mathrm{p} \simeq 1.4{\times}10^3~\mathrm{kg~m^{-3}}$. The particle image observed by a stereomicroscope (LEICA Z16 APO) is shown in Fig.~\ref{fig:setup}(b). The particles are roughly ellipsoidal, with some variation in shape from grain to grain.

As the suspending fluid, aqueous solutions of sodium polytungstate (SPT) were used. After preparing suspension at a designated volume fraction, the prepared suspension was poured into the container of inner dimensions $60{\times}29{\times}12$~mm (width $\times$ depth $\times$ height). The density of the SPT solution $\rho_\mathrm{L}$ was set to $1.0,\ 1.1,\ 1.2,\ 1.3$, or $1.4{\times}10^3~\mathrm{kg~m^{-3}}$, to produce five different density contrasts ($\rho_\mathrm{p} - \rho_\mathrm{L}$). The initial volume fraction $\phi_0$ was varied from 0.30 to 0.50 in steps of 0.02. Each parameter combination was repeated five times to check reproducibility.

For each run, the suspension was mixed quickly by hand to achieve an approximately uniform particle distribution throughout the container, and then placed under the OCT probe. Measurements started immediately after mixing, so that the initial instance ($t=0$) corresponds to a nominally homogeneous suspension at the initial volume fraction $\phi_0$.

For comparison, kaolin suspensions were also imaged with the OCT system (Fig.~\ref{fig:setup}(c)). Kaolin has a much smaller particle size (${\sim}5~\mathrm{\mu m}$), comparable with the OCT resolution. This comparison illustrates how particle size affects the OCT signal, as discussed in Sec.~\ref{sec:results:images}.

\section{Results}
\label{sec:results}

\subsection{OCT images: potato starch versus kaolin}
\label{sec:results:images}

Figure~\ref{fig:setup}(d) shows a representative OCT image of a potato-starch suspension. Individual particles appear as bright dots scattered throughout the image. This is because the particle diameter (${\sim}20~\mathrm{\mu m}$) exceeds the OCT pixel size (${\sim}1.9~\mathrm{\mu m/pixel}$), so each particle is resolved as a distinct scatterer. A bright horizontal line near the top of the image corresponds to the reflection from the liquid surface. Just below it, a nearly dark region is visible. This is the supernatant layer, where particle concentration is too low to produce significant backscattering. Further below, the bright-dot cluster represents the particle-rich suspension layer. Dark regions at greater depths indicate that the OCT probe light cannot penetrate further due to strong multiple scattering in the dense suspension, even though these regions are also filled with potato-starch particles. Thus, only particles within a thin layer near the top of the sedimenting suspension can be visualized with the OCT system used in this study.

In contrast, kaolin particles produce a qualitatively different image (Fig.~\ref{fig:setup}(e)). The image shows a uniform speckle pattern rather than individual dots. Kaolin particles are roughly $5~\mathrm{\mu m}$ in diameter, comparable with the OCT resolution. 
When the scatterer size is smaller than the coherence length of the light source, multiple particles contribute coherently to each image pixel, resulting in a speckle pattern~\citep{Almasian:2017}.
This comparison clearly shows that reliable particle imaging using the current OCT system requires the particle size to sufficiently exceed the instrument resolution. Potato starch meets this condition, while kaolin does not.

\subsection{Time evolution of sedimentation}
\label{sec:results:timeevo}

\begin{figure*}
  \centering
  \includegraphics[width=\linewidth]{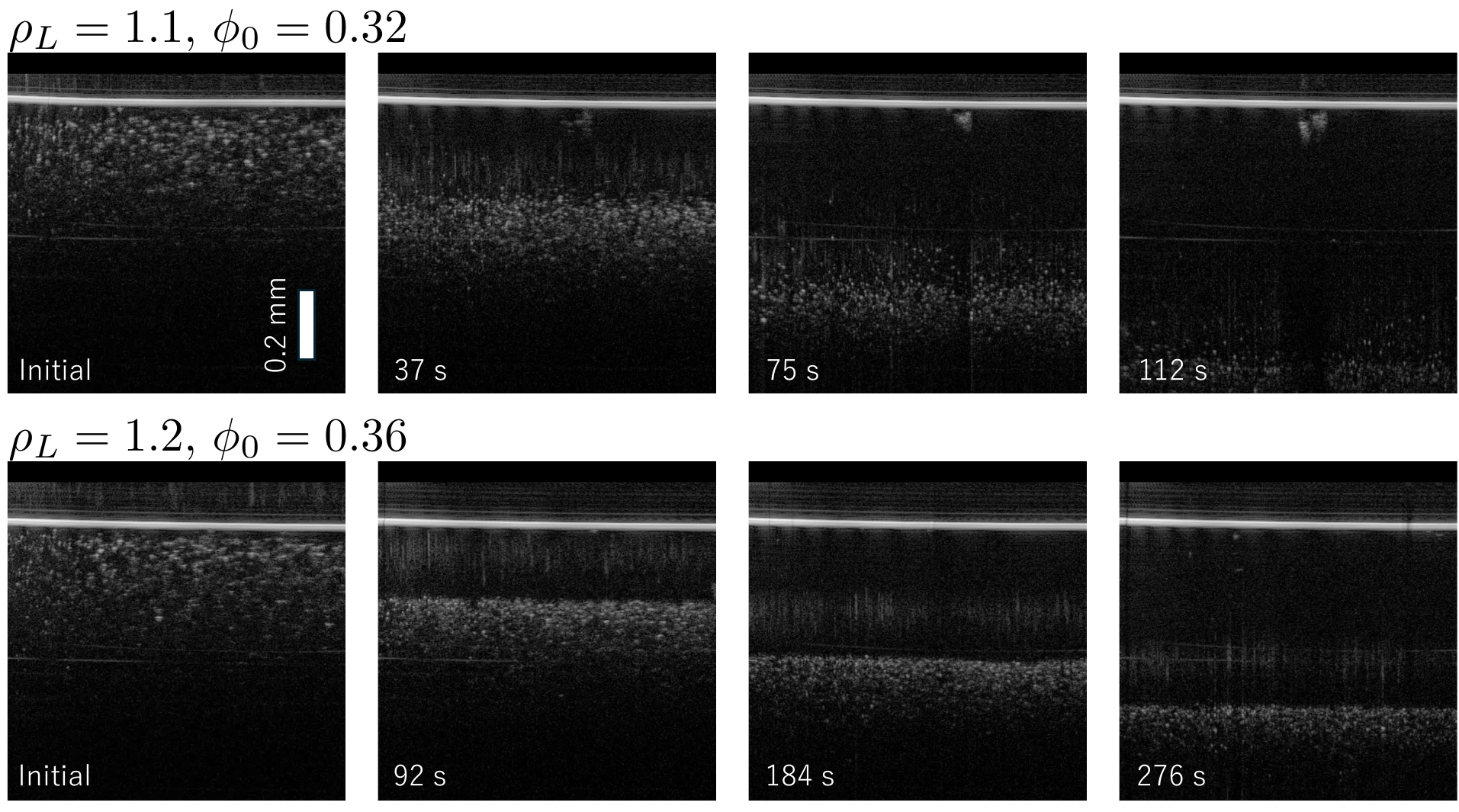}
  \caption{Time series of OCT snapshots for (top)~$\rho_\mathrm{L} = 1.1{\times}10^3~\mathrm{kg~m^{-3}}$ and $\phi_0 = 0.32$, and (bottom)~$\rho_\mathrm{L} = 1.2{\times}10^3~\mathrm{kg~m^{-3}}$ and $\phi_0 = 0.36$. The particle cluster moves downward as the supernatant layer grows from above. Time $t$ after mixing the suspension is labeled at the left bottom of each snapshot. By tracking the centroid of the particle cluster over time, the sedimentation velocity can be measured.}
  \label{fig:snapshots}
\end{figure*}

Figure~\ref{fig:snapshots} shows a series of OCT snapshots taken at different times for two representative conditions ($\rho_\mathrm{L} = 1.1{\times}10^3~\mathrm{kg~m^{-3}}$, $\phi_0 = 0.32$ and $\rho_\mathrm{L} = 1.2{\times}10^3~\mathrm{kg~m^{-3}}$, $\phi_0 = 0.36$). At early times, the bright-dot cluster fills the upper part of the images. As time proceeds, the cluster moves downward and the supernatant layer grows from above. The descent of the particle cluster and the growth of the supernatant layer are both clearly visible in the image sequence.

\subsection{Image analysis}
\label{sec:results:analysis}

The OCT images were binarized and analyzed to extract two quantities: the centroid depth of the particle layer $\langle Z \rangle$, and the position of the lower boundary of the supernatant layer $Z_\mathrm{sup}$.

Because gravity-driven sedimentation primarily produces motion in the $z$ direction, the present analysis focuses on tracking the depth-wise (z) position of the particle layer rather than resolving the full three-dimensional velocity field. As noted in Sec.~\ref{sec:exp:OCT}, the OCT image captures a two-dimensional $x$--$z$ slice of the suspension, similar in spirit to a laser-sheet cross-section in PIV. The analysis does not track individual particle trajectories. Instead, the brightness is summed along $x$ at each depth $z$, and only the resulting depth-wise centroid is used. Particles continuously enter and leave the imaged $x$--$z$ slice along $y$. This exchange is statistically balanced under the quasi-steady sedimentation condition. As a result, the centroid position $\langle Z \rangle(t)$ remains a meaningful measure of the bulk descent rate, even though the underlying flow is three-dimensional. Because the suspension is laterally uniform on average, the binarized image brightness (0 or 1) was first averaged horizontally at each depth $z$ and each time frame $t$. This gives a one-dimensional brightness profile $I(z, t)$. From this profile, $\langle Z \rangle (t)$ was computed as the intensity-weighted centroid:
\begin{equation}
  \langle Z \rangle(t) = \frac{\sum_z z \, I(z, t)}{\sum_z I(z, t)}.
  \label{eq:centroid}
\end{equation}
The sedimentation velocity $V(t)$ was obtained by numerical differentiation of $\langle Z \rangle(t)$ with respect to time $t$. In practice, $V(t)$ was computed by linearly fitting $\langle Z \rangle$ within a $\pm 15$~s window centered on each time step. The size of the fitting window does not qualitatively affect the resulting $V(t)$. Note that $V(t) = d\langle Z\rangle/dt$ measures the rate at which the imaged particle layer descends rather than the velocity of any individual particle. 

Then, the supernatant boundary $Z_\mathrm{sup}(t)$ was identified as the depth at which the brightness profile drops to a threshold value. In practice, we used the depth at which $I(z,t)$ falls to 80\% of its peak value. The exact threshold value does not affect the results qualitatively. $Z_\mathrm{sup}(t)$ is later used to compute the time-evolving effective volume fraction $\phi_\mathrm{eff}(t)$ (Sec.~\ref{sec:discussion:phieff}). The bright horizontal line at the top of each image corresponds to the reflection from the liquid surface and was used as a fixed reference position.

Note that for $\rho_\mathrm{L} = 1.4{\times}10^3~\mathrm{kg~m^{-3}}$, gravity and buoyancy on the particles nearly cancel, and clear sedimentation was not observed over the measurement period. This condition was therefore excluded from further analysis.

\subsection{Centroid tracking and supernatant boundary}
\label{sec:results:tracking}

\begin{figure}
  \centering
  \includegraphics[width=\linewidth]{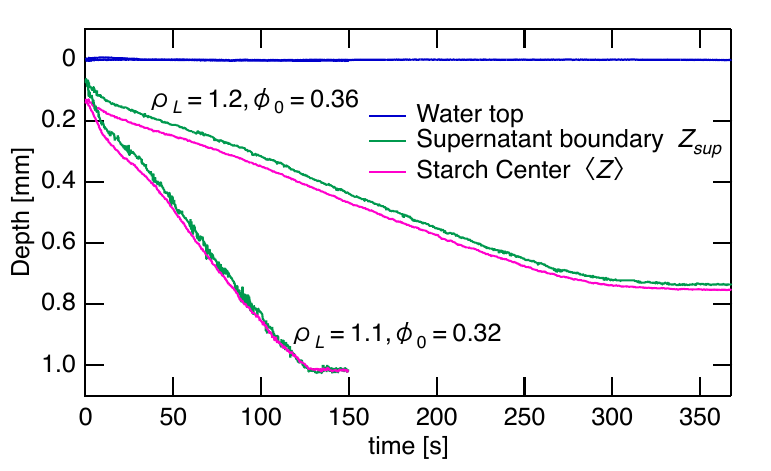}
  \caption{Time evolution of the water surface, supernatant boundary $Z_\mathrm{sup}$, and the centroid position of the particle layer $\langle Z \rangle$ for two representative conditions corresponding to the data shown in Fig.~\ref{fig:snapshots}.}
  \label{fig:tracks}
\end{figure}

Figure~\ref{fig:tracks} shows the time evolution of water surface level, $\langle Z \rangle (t)$, and $Z_\mathrm{sup}(t)$ for two representative conditions. Both $\langle Z \rangle$ and $Z_\mathrm{sup}$ show basically monotonic variation with time. This reflects the steady downward motion of the particle layer and the simultaneous growth of the supernatant from the top.

For most conditions, $\langle Z \rangle(t)$ varies at a roughly constant rate. The instantaneous settling velocity $V(t)$ was measured from $\langle Z \rangle(t)$. In some conditions, $\langle Z \rangle(t)$ develops a shoulder-like shape indicating a non-monotonic variation of $V(t)$. A hint of this behavior is visible around $t\simeq 20$~s for the $\rho_\mathrm{L}=1.1{\times}10^3$~kg~m$^{-3}$, $\phi_0=0.32$. This oscillatory behavior is more pronounced under certain conditions. A detailed analysis of these fluctuations is left for future work. In what follows we focus on the dominant sedimentation process.

\section{Analysis and Discussion}
\label{sec:discussion}

\subsection{Effective volume fraction}
\label{sec:discussion:phieff}

As sedimentation proceeds, a supernatant layer grows from the top. The effective particle volume fraction in the remaining suspension region therefore increases over time. We define the effective volume fraction $\phi_\mathrm{eff}(t)$ as
\begin{equation}
  \phi_\mathrm{eff}(t) = \frac{\mathcal{V}_\mathrm{p}}{\mathcal{V}_\mathrm{total} - \mathcal{V}_\mathrm{sup}(t)},
  \label{eq:phieff}
\end{equation}
where $\mathcal{V}_\mathrm{p}$ is the total volume of potato-starch particles (constant), $\mathcal{V}_\mathrm{total}$ is the total suspension volume, and $\mathcal{V}_\mathrm{sup}(t)$ is the volume of the supernatant layer at time $t$. The supernatant volume is estimated from $Z_\mathrm{sup}(t)$ and the known cross-sectional area of 
the container. As the supernatant grows, $\phi_\mathrm{eff}$ increases accordingly.
This definition assumes the supernatant layer is particle-free, which is consistent with the sharp drop in OCT intensity across $Z_\mathrm{sup}(t)$. 
As Fig.~\ref{fig:tracks} shows, $Z_\mathrm{sup}$ remains close to $\langle Z \rangle$ throughout each run. This means that the imaged particle-rich region stays thin and well-defined.

In this study, $V(t)$ and $\phi_\mathrm{eff}(t)$ are obtained simultaneously at each time step. The whole sedimentation process therefore traces out a continuous trajectory in the $V$--$\phi_\mathrm{eff}$ plane. In contrast, conventional batch measurements typically report only a single run-averaged data point per experiment, even though continuous tracking of the supernatant boundary is in principle possible. A single OCT run here spans a substantial range of $\phi_\mathrm{eff}$ within one continuous trajectory. This provides a more complete experimental test of hindered-settling models. This definition assumes that no significant particle deposit has yet formed at the container bottom, which is supported by the fact that all the data analyzed here correspond to times before $\langle Z\rangle$ approaches the bottom of the container.

\subsection{Sedimentation models}
\label{sec:discussion:models}

We compare the measured $V(t)$ versus $\phi_\mathrm{eff}(t)$ with two standard models.

The first is the Krieger--Dougherty (K-D) model~\citep{Krieger:1959}. The starting point is the Stokes terminal velocity of a single isolated sphere in a fluid of viscosity $\eta_0$,
\begin{equation}
  V_\mathrm{Stokes}  = \frac{(\rho_\mathrm{p} - \rho_\mathrm{L})\,g\,D_\mathrm{p}^2}{18\,\eta_0}.
  \label{eq:stokes}
\end{equation}
Here, $g=9.8$~m~s$^{-2}$ is gravitational acceleration. This expression is strictly valid only for a single sphere in an unbounded, quiescent fluid. The models compared in this section correct this idealized expression for the effects of finite particle concentration, and deviations arising from particle shape and polydispersity are also reflected in the values of the resulting fitting parameters (Sec.~\ref{sec:discussion:comparison}). To compute $V_\mathrm{Stokes}$, we use $D_\mathrm{p} =20$~$\mu$m and $\eta_0=1.0$~mPa~s (the viscosity of pure water) together with the appropriate density contrast $(\rho_\mathrm{p}-\rho_\mathrm{L})$. The viscosity of SPT solutions increases with density. Using the water value therefore overestimates $V_\mathrm{Stokes}$ at high $\rho_\mathrm{L}$. The resulting bias is modest, however, because the density range employed here ($\rho_\mathrm{L}\lesssim 1.3{\times}10^3$~kg~m$^{-3}$) does not produce a large viscosity increase. The implications of this approximation are discussed in Sec.~\ref{sec:discussion:comparison}. The effective viscosity $\eta_\mathrm{eff}$ proposed by the K-D relation is written as,
\begin{equation}
  \eta_\mathrm{eff} = \eta_0 \left(1 - \frac{\phi_\mathrm{eff}}{\phi_\mathrm{max}} \right)^{-[\eta]\phi_\mathrm{max}}.
  \label{eq:KD_eta}
\end{equation}
Here, $\phi_\mathrm{max}$ is the maximum volume fraction and $[\eta]$ is a parameter characterizing nonlinear viscosity increase. Both $\phi_\mathrm{max}$ and $[\eta]$ are treated as free fitting parameters in this work. We assume that a settling particle experiences the suspension as an effective Newtonian fluid of viscosity $\eta_\mathrm{eff}$. Then, the terminal velocity based on the K-D relation $V_\mathrm{KD}$ is 
\begin{equation}
  V_\mathrm{KD} =V_\mathrm{Stokes}\frac{\eta_0}{\eta_\mathrm{eff}}.
  \label{eq:KD}
\end{equation}

The second model is the Richardson--Zaki (R-Z) law~\citep{Richardson:1997}. Based on the R-Z law, the sedimentation velocity $V_\mathrm{RZ}$ is written as,
\begin{equation}
  \frac{V_\mathrm{RZ}}{V_\mathrm{Stokes}} = (1 - \phi_\mathrm{eff})^n,
  \label{eq:RZ}
\end{equation}
where $n$ is an exponent depending on the particle Reynolds number. The dependence of $n$ on the Reynolds number has been examined systematically for various solid-liquid systems~\citep{Garside:1977}. In the Stokes regime ($\mathrm{Re}_\mathrm{p} \ll 1$), $n \simeq 4.65$~\citep{Richardson:1997}. The particle Reynolds number in our experiments is estimated as $\mathrm{Re}_\mathrm{p} \sim 10^{-4}$--$10^{-3}$, confirming that we are safely in the Stokes regime.

\subsection{Comparison with data}
\label{sec:discussion:comparison}
\begin{figure}
  \centering
  \includegraphics[width=\linewidth]{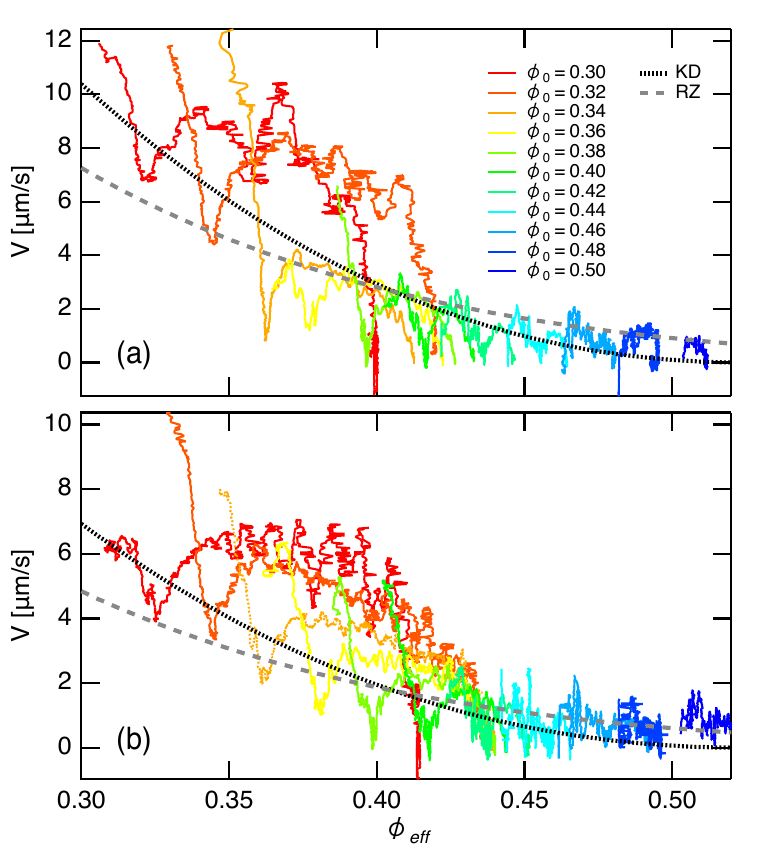}
  \caption{Settling velocity $V$ versus effective volume fraction $\phi_\mathrm{eff}$ for two representative conditions: $\rho_\mathrm{L}=1.1{\times}10^3$ or $1.2{\times}10^3$~kg~m$^{-3}$. Black dotted and gray dashed curves show the K-D model (Eq.~\ref{eq:KD}) and the R-Z law (Eq.~\ref{eq:RZ}), respectively. The parameter values in these models are determined by the fitting shown in Fig.~\ref{fig:all}. The color code is identical to that of Fig.~\ref{fig:tracks}.}
  \label{fig:single}
\end{figure}

We determine the model parameters ($\phi_\mathrm{max}$ and $[\eta]$ for the K-D model and $n$ for the R-Z law) by a global least-squares fit to the full data set. For clarity, however, we first show representative data examples. 
Figure~\ref{fig:single} shows $V$ as a function of $\phi_\mathrm{eff}$ for two representative conditions, with the K-D and R-Z predictions overlaid. Each data point corresponds to one time frame and each curve indicates one experimental run. The trajectory therefore moves from lower $\phi_\mathrm{eff}$ at early times to higher $\phi_\mathrm{eff}$ at late times. As seen in Fig.~\ref{fig:single}, both curves (K-D and R-Z) track the data while the fitting quality is slightly better for the K-D model. 

\begin{figure}
  \centering
  \includegraphics[width=\linewidth]{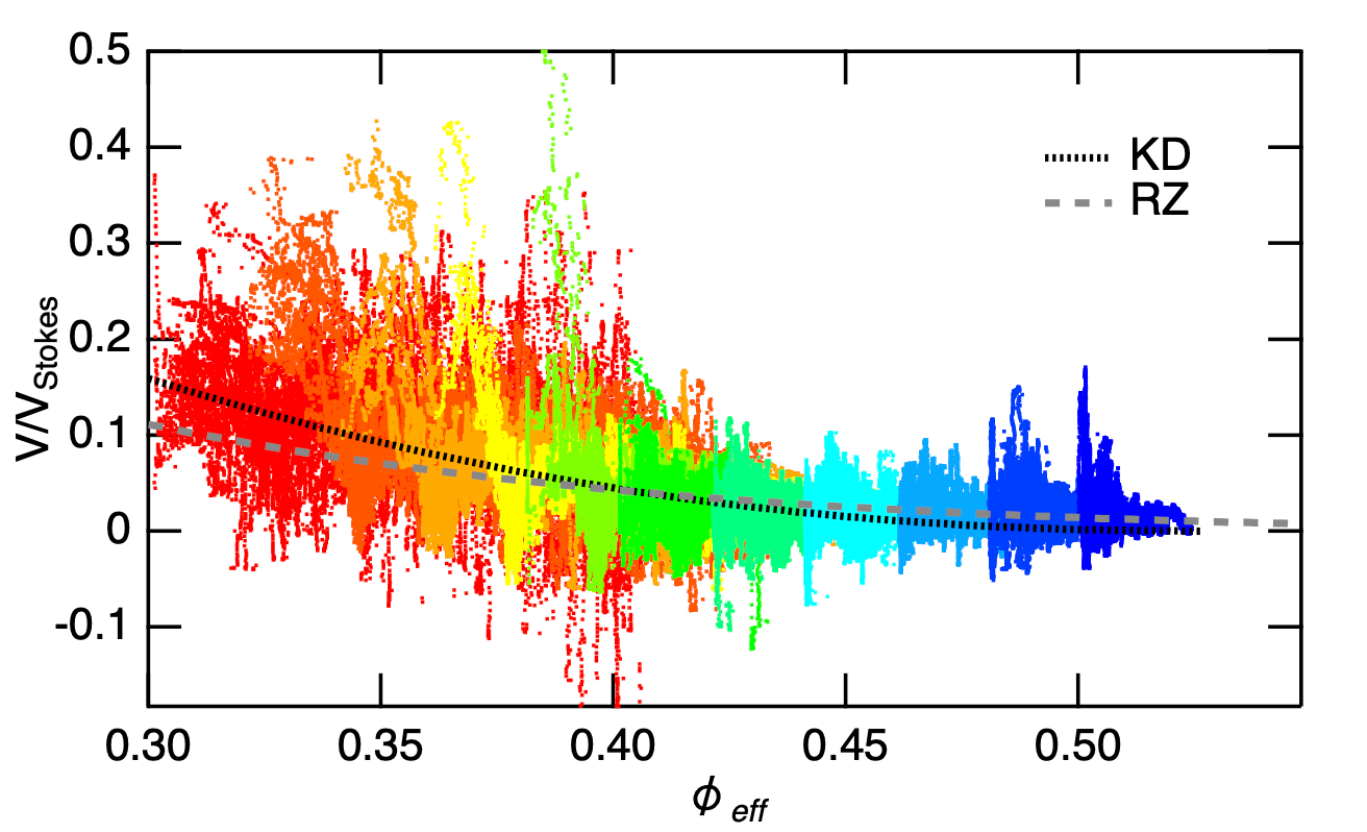}
  \caption{Normalized settling velocity $V/V_\mathrm{Stokes}$ versus $\phi_\mathrm{eff}$ for all experimental conditions. Black dotted and gray dashed curves show the K-D model and R-Z law, respectively. The parameter values $\phi_\mathrm{max}=0.53$ and $[\eta]=4.1$ for K-D model, and $n=6.2$ for R-Z model are determined by the least-square fitting.}
  \label{fig:all}
\end{figure}

To evaluate all conditions together, $V$ is normalized by $V_\mathrm{Stokes}$. Figure~\ref{fig:all} shows data from all experimental conditions plotted as $V/V_\mathrm{Stokes}$ versus $\phi_\mathrm{eff}$. Data from different $\phi_0$ and $\rho_\mathrm{L}$ conditions collapse onto a common trend, indicating that $\phi_\mathrm{eff}$ is the relevant control variable. The overall trend is consistent with both K-D and R-Z predictions over $\phi_\mathrm{eff} \simeq 0.30$--$0.52$. The fitted parameters are $\phi_\mathrm{max}=0.53$, $[\eta]=4.1$, and $n=6.2$. Since the K-D model has more fitting parameters, it fits the data slightly better. The estimated $\phi_\mathrm{max}$ is in a physically reasonable range. This agreement confirms that OCT-based tracking yields physically reasonable hindered-settling behavior, consistent with established models.

The fitted values $[\eta]=4.1$ and $n=6.2$ are larger than the canonical values for spheres (the Einstein value $[\eta]=2.5$ and the Stokes-regime value $n\simeq 4.65$). This is consistent with the non-spherical shape and polydispersity of potato-starch grains~\citep{Pabst:2006,Mueller:2009}. We note also that $\eta_0=1.0$~mPa~s was used for all conditions, which slightly overestimates $V_\mathrm{Stokes}$ at high $\rho_\mathrm{L}$. A proper viscosity correction would shift $V/V_\mathrm{Stokes}$ upward at high $\rho_\mathrm{L}$ and would slightly reduce the fitted parameters toward their canonical values. In any case, the collapse of all conditions onto a common trend (Fig.~\ref{fig:all}) indicates that the $\phi_\mathrm{eff}$ dependence dominates over this correction within the present scatter. A more accurate analysis with measured SPT viscosities is left for future work. 

Two aspects of these results are worth separating. The functional form of both K-D and R-Z captures the overall data trend over all conditions. The parameter values, while larger than the hard-sphere canonical values, are qualitatively consistent with the non-spherical shape and size polydispersity of potato-starch grains. No additional mechanism related to DST or impact solidification is needed to describe the quasi-static sedimentation data. This conclusion is supported by the separation of timescales. The characteristic shear rate associated with sedimentation, $\dot{\gamma}_\mathrm{sed} \sim V/D_\mathrm{p} \lesssim 1$~s$^{-1}$, is well below the DST onset shear rates reported for potato-starch suspensions ($\dot{\gamma}_\mathrm{c} \simeq 10$~s$^{-1}$ or higher~\citep{Brown:2014}).

The two models are difficult to conclusively distinguish at the present scatter level. Both give a similar functional form over this concentration range. This is not specific to our data. \citet{Hamid:2013} showed through numerical simulations that K-D and R-Z give comparable predictions over a wide range of volume fractions, and that discriminating between them requires very low scatter in the data. Recent simulations by \citet{Kundu:2025} also confirm that both models align well with computed settling velocities up to high volume fractions. In summary, the master curve of $V/V_\mathrm{Stokes}$ versus $\phi_\mathrm{eff}$ (Fig.~\ref{fig:all}) implies that potato-starch sedimentation falls within the ordinary hindered-settling regime, with no need to invoke DST or impact-solidification responses associated with high strain rates.
A definitive discrimination between the K-D and R-Z functional forms is beyond the scope of the present work. 

\subsection{On the scatter in the data}
\label{sec:discussion:scatter}

The data show more scatter than we would expect from measurement noise alone, and several physical factors probably play a role. The most likely is the narrowness of the OCT observation window (${\sim}1~\mathrm{mm}$ laterally), which means that the centroid reflects only a small cross-section of the suspension. Local fluctuations in particle concentration are known to be large in sedimenting suspensions, even under idealized conditions~\citep{Caflisch:1985,Segre:1997,Cunha:2002}. In addition, residual convective motion from the initial mixing may persist. This may cause systematic drift in the early-time data. Finally, potato-starch particles are not perfect spheres and have a distribution of sizes, both of which will broaden the distribution of individual settling velocities. We cannot fully separate these contributions from the present data set, but the scatter is broadly consistent with the velocity fluctuations reported for non-Brownian suspensions in the literature~\citep{Nicolai:1995,Guazzelli:2011}. In some conditions, the centroid trajectory shows a reproducible shoulder-like deviation rather than purely random scatter (Sec.~\ref{sec:results:tracking}). This suggests that part of the fluctuation may reflect an underlying systematic behavior rather than noise alone. While these fluctuations broaden the data scatter, they do not alter the central observation that $V/V_\mathrm{Stokes}$ collapses onto a $\phi_\mathrm{eff}$-dependent master curve consistent with the K-D and R-Z forms (Fig.~\ref{fig:all}). 
A conventional batch-averaged measurement reports a single value of $V$ per run. Such averaging would likely make the scatter look smaller. We deliberately avoid this averaging here. The scatter reported in this study becomes visible precisely because the measurement is time-resolved. A detailed physical account of this fluctuation is left for future work. The present study instead prioritizes establishing OCT as a method for direct, time-resolved observation of dense sedimentation. This scatter itself is an interesting problem. Its emergence shows that OCT adds a new experimental approach to the study of concentrated suspension sedimentation.

\section{Conclusion}
\label{sec:conclusion}

We have used optical coherence tomography (OCT) to obtain time-resolved, simultaneous measurements of the settling velocity $V(t)$ and the effective volume fraction $\phi_\mathrm{eff}(t)$ in dense potato-starch suspensions. Sedimentation was measured for initial volume fractions $\phi_0 = 0.30$--$0.50$ (yielding $\phi_\mathrm{eff}$ up to $\sim 0.52$) and solvent densities $\rho_\mathrm{L}$ from $1.0$ to $1.3{\times}10^3$~kg~m$^{-3}$ (the $\rho_\mathrm{L}=1.4{\times}10^3$~kg~m$^{-3}$ case showed no clear sedimentation and was excluded).

First, and most importantly, this study demonstrates that OCT can directly resolve the internal dynamics of a dense, optically opaque suspension during sedimentation, a regime that has been inaccessible to existing optical techniques. OCT can resolve individual potato-starch particles ($D_\mathrm{p} \simeq 20~\mathrm{\mu m}$) as distinct scatterers, which is challenging for conventional optical microscopy. Kaolin particles, which are much smaller, produced only a speckle pattern and could not be tracked individually. The particle size relative to the OCT resolution is the key factor. This capability itself constitutes the principal contribution of the present work, independent of the specific functional forms used to describe the data below.

Second, by tracking both the particle centroid $\langle Z \rangle(t)$ and the supernatant boundary $Z_\mathrm{sup}(t)$ simultaneously, we obtained the settling velocity $V(t)$ and the effective volume fraction $\phi_\mathrm{eff}(t)$ as continuous functions of time. This kind of time-resolved, simultaneous measurement of $V$ and $\phi_\mathrm{eff}$ during a single sedimentation run is not achievable in conventional batch experiments.

Third, when the normalized settling velocity $V/V_\mathrm{Stokes}$ is plotted against $\phi_\mathrm{eff}$, the data from all conditions collapse onto a common trend. Both the K-D model and the R-Z law describe this trend reasonably well over the range $\phi_\mathrm{eff} \simeq 0.30$--$0.52$. Although the fitted value $\phi_\mathrm{max}=0.53$ is reasonable, the other fitted parameter values $[\eta]=4.1$ and $n=6.2$ exceed the canonical hard-sphere values. However, this result is consistent with the non-spherical shape and polydispersity of potato-starch grains. Beyond these shape-related deviations, no signature specific to DST or impact solidification is seen in the quasi-static sedimentation data. In other words, gravity-driven sedimentation in this rheologically complex system follows the same empirical forms as ordinary non-Brownian suspensions.

The scatter in the velocity data is large, and we attribute this partly to the narrow OCT observation window and partly to hydrodynamic velocity fluctuations inherent to dense sedimenting suspensions. The size distribution of sedimenting particles likely contributes as well. A quantitative characterization of these large fluctuations remains a topic for future work.

\section*{Acknowledgements}
This work was supported by JSPS KAKENHI Grant Numbers: JP23H04134 and JP25K22010, and JST ERATO Grant Number JPMJER2401.

\section*{Data Availability}
The data that support the findings of this study are available upon reasonable request.

\section*{Declaration of competing interests}
The authors declare that they have no known competing financial interests or personal relationships that could have appeared to influence the work reported in this paper.

\section*{CRediT authorship contribution statement}
\textbf{Tasuku Saiki:} Investigation, Data curation, Formal analysis, Visualization. 
\textbf{Hiroaki Katsuragi:} Conceptualization, Supervision, Writing -- original draft, Writing -- review \& editing, Funding acquisition.

\bibliographystyle{elsarticle-harv} 
\bibliography{StarchSediment}

\begin{thebibliography}{28}
\expandafter\ifx\csname natexlab\endcsname\relax\def\natexlab#1{#1}\fi
\providecommand{\url}[1]{\texttt{#1}}
\providecommand{\href}[2]{#2}
\providecommand{\path}[1]{#1}
\providecommand{\DOIprefix}{doi:}
\providecommand{\ArXivprefix}{arXiv:}
\providecommand{\URLprefix}{URL: }
\providecommand{\Pubmedprefix}{pmid:}
\providecommand{\doi}[1]{\href{http://dx.doi.org/#1}{\path{#1}}}
\providecommand{\Pubmed}[1]{\href{pmid:#1}{\path{#1}}}
\providecommand{\bibinfo}[2]{#2}
\ifx\xfnm\relax \def\xfnm[#1]{\unskip,\space#1}\fi
\bibitem[{Al-Naafa and Selim(1992)}]{AlNaafa:1992}
\bibinfo{author}{Al-Naafa, M.A.}, \bibinfo{author}{Selim, M.S.},
  \bibinfo{year}{1992}.
\newblock \bibinfo{title}{{Sedimentation of monodisperse and bidisperse
  hard-sphere colloidal suspensions}}.
\newblock \bibinfo{journal}{AIChE Journal} \bibinfo{volume}{38},
  \bibinfo{pages}{1618--1630}.
\newblock \DOIprefix\doi{10.1002/aic.690381012}.
\bibitem[{Almasian et~al.(2017)Almasian, Leeuwen and Faber}]{Almasian:2017}
\bibinfo{author}{Almasian, M.}, \bibinfo{author}{Leeuwen, T.G.v.},
  \bibinfo{author}{Faber, D.J.}, \bibinfo{year}{2017}.
\newblock \bibinfo{title}{{OCT Amplitude and Speckle Statistics of Discrete
  Random Media}}.
\newblock \bibinfo{journal}{Scientific Reports} \bibinfo{volume}{7},
  \bibinfo{pages}{14873}.
\newblock \DOIprefix\doi{10.1038/s41598-017-14115-3}.
\bibitem[{Amini et~al.(2025)Amini, Wittig, Saoncella, Tammisola, Lundell and
  Bagheri}]{Amini:2025}
\bibinfo{author}{Amini, K.}, \bibinfo{author}{Wittig, C.},
  \bibinfo{author}{Saoncella, S.}, \bibinfo{author}{Tammisola, O.},
  \bibinfo{author}{Lundell, F.}, \bibinfo{author}{Bagheri, S.},
  \bibinfo{year}{2025}.
\newblock \bibinfo{title}{{Optical coherence tomography in soft matter}}.
\newblock \bibinfo{journal}{Soft Matter} \bibinfo{volume}{21},
  \bibinfo{pages}{3425--3442}.
\newblock \DOIprefix\doi{10.1039/d4sm01537a}.
\bibitem[{Batchelor(1972)}]{Batchelor:1972}
\bibinfo{author}{Batchelor, G.K.}, \bibinfo{year}{1972}.
\newblock \bibinfo{title}{{Sedimentation in a dilute dispersion of spheres}}.
\newblock \bibinfo{journal}{Journal of Fluid Mechanics} \bibinfo{volume}{52},
  \bibinfo{pages}{245--268}.
\newblock \DOIprefix\doi{10.1017/s0022112072001399}.
\bibitem[{Brown and Jaeger(2014)}]{Brown:2014}
\bibinfo{author}{Brown, E.}, \bibinfo{author}{Jaeger, H.M.},
  \bibinfo{year}{2014}.
\newblock \bibinfo{title}{{Shear thickening in concentrated suspensions:
  phenomenology, mechanisms and relations to jamming}}.
\newblock \bibinfo{journal}{Reports on Progress in Physics}
  \bibinfo{volume}{77}, \bibinfo{pages}{046602}.
\newblock \DOIprefix\doi{10.1088/0034-4885/77/4/046602}.
\bibitem[{Buchsbaum et~al.(2015)Buchsbaum, Egger, Burzic, Koepplmayr, Aigner,
  Miethlinger and Leitner}]{Buchsbaum:2015}
\bibinfo{author}{Buchsbaum, A.}, \bibinfo{author}{Egger, M.},
  \bibinfo{author}{Burzic, I.}, \bibinfo{author}{Koepplmayr, T.},
  \bibinfo{author}{Aigner, M.}, \bibinfo{author}{Miethlinger, J.},
  \bibinfo{author}{Leitner, M.}, \bibinfo{year}{2015}.
\newblock \bibinfo{title}{{Optical coherence tomography based particle image
  velocimetry (OCT-PIV) of polymer flows}}.
\newblock \bibinfo{journal}{Optics and Lasers in Engineering}
  \bibinfo{volume}{69}, \bibinfo{pages}{40--48}.
\newblock \DOIprefix\doi{10.1016/j.optlaseng.2015.02.003}.
\bibitem[{Caflisch and Luke(1985)}]{Caflisch:1985}
\bibinfo{author}{Caflisch, R.E.}, \bibinfo{author}{Luke, J.H.C.},
  \bibinfo{year}{1985}.
\newblock \bibinfo{title}{{Variance in the sedimentation speed of a
  suspension}}.
\newblock \bibinfo{journal}{Physics of Fluids} \bibinfo{volume}{28},
  \bibinfo{pages}{759--760}.
\newblock \DOIprefix\doi{10.1063/1.865095}.
\bibitem[{Cunha et~al.(2002)Cunha, Abade, Sousa and Hinch}]{Cunha:2002}
\bibinfo{author}{Cunha, F.R.}, \bibinfo{author}{Abade, G.C.},
  \bibinfo{author}{Sousa, A.J.}, \bibinfo{author}{Hinch, E.J.},
  \bibinfo{year}{2002}.
\newblock \bibinfo{title}{{Modeling and Direct Simulation of Velocity
  Fluctuations and Particle-Velocity Correlations in Sedimentation}}.
\newblock \bibinfo{journal}{Journal of Fluids Engineering}
  \bibinfo{volume}{124}, \bibinfo{pages}{957--968}.
\newblock \DOIprefix\doi{10.1115/1.1502665}.
\bibitem[{Davis and Acrivos(1985)}]{Davis:1985}
\bibinfo{author}{Davis, R.H.}, \bibinfo{author}{Acrivos, A.},
  \bibinfo{year}{1985}.
\newblock \bibinfo{title}{{Sedimentation of Noncolloidal Particles at Low
  Reynolds Numbers}}.
\newblock \bibinfo{journal}{Annual Review of Fluid Mechanics}
  \bibinfo{volume}{17}, \bibinfo{pages}{91--118}.
\newblock \DOIprefix\doi{10.1146/annurev.fl.17.010185.000515}.
\bibitem[{Egawa and Katsuragi(2019)}]{Egawa:2019}
\bibinfo{author}{Egawa, K.}, \bibinfo{author}{Katsuragi, H.},
  \bibinfo{year}{2019}.
\newblock \bibinfo{title}{{Bouncing of a projectile impacting a dense
  potato-starch suspension layer}}.
\newblock \bibinfo{journal}{Physics of Fluids} \bibinfo{volume}{31},
  \bibinfo{pages}{053304}.
\newblock \DOIprefix\doi{10.1063/1.5095678}.
\bibitem[{Fercher et~al.(2003)Fercher, Drexler, Hitzenberger and
  Lasser}]{Fercher:2003}
\bibinfo{author}{Fercher, A.F.}, \bibinfo{author}{Drexler, W.},
  \bibinfo{author}{Hitzenberger, C.K.}, \bibinfo{author}{Lasser, T.},
  \bibinfo{year}{2003}.
\newblock \bibinfo{title}{{Optical coherence tomography --- principles and
  applications}}.
\newblock \bibinfo{journal}{Reports on Progress in Physics}
  \bibinfo{volume}{66}, \bibinfo{pages}{239--303}.
\newblock \DOIprefix\doi{10.1088/0034-4885/66/2/204}.
\bibitem[{Garside and Al-Dibouni(1977)}]{Garside:1977}
\bibinfo{author}{Garside, J.}, \bibinfo{author}{Al-Dibouni, M.R.},
  \bibinfo{year}{1977}.
\newblock \bibinfo{title}{{Velocity-Voidage Relationships for Fluidization and
  Sedimentation in Solid-Liquid Systems}}.
\newblock \bibinfo{journal}{Industrial \& Engineering Chemistry Process Design
  and Development} \bibinfo{volume}{16}, \bibinfo{pages}{206--214}.
\newblock \DOIprefix\doi{10.1021/i260062a008}.
\bibitem[{Guazzelli and Hinch(2011)}]{Guazzelli:2011}
\bibinfo{author}{Guazzelli, {\'E}.}, \bibinfo{author}{Hinch, J.},
  \bibinfo{year}{2011}.
\newblock \bibinfo{title}{{Fluctuations and Instability in Sedimentation}}.
\newblock \bibinfo{journal}{Annual Review of Fluid Mechanics}
  \bibinfo{volume}{43}, \bibinfo{pages}{97--116}.
\newblock \DOIprefix\doi{10.1146/annurev-fluid-122109-160736}.
\bibitem[{Hamid et~al.(2013)Hamid, Molina and Yamamoto}]{Hamid:2013}
\bibinfo{author}{Hamid, A.}, \bibinfo{author}{Molina, J.J.},
  \bibinfo{author}{Yamamoto, R.}, \bibinfo{year}{2013}.
\newblock \bibinfo{title}{{Sedimentation of non-Brownian spheres at high volume
  fractions}}.
\newblock \bibinfo{journal}{Soft Matter} \bibinfo{volume}{9},
  \bibinfo{pages}{10056--10068}.
\newblock \DOIprefix\doi{10.1039/c3sm50748c}.
\bibitem[{Huang et~al.(1991)Huang, Swanson, Lin, Schuman, Stinson, Chang, Hee,
  Flotte, Gregory, Puliafito and Fujimoto}]{Huang:1991}
\bibinfo{author}{Huang, D.}, \bibinfo{author}{Swanson, E.A.},
  \bibinfo{author}{Lin, C.P.}, \bibinfo{author}{Schuman, J.S.},
  \bibinfo{author}{Stinson, W.G.}, \bibinfo{author}{Chang, W.},
  \bibinfo{author}{Hee, M.R.}, \bibinfo{author}{Flotte, T.},
  \bibinfo{author}{Gregory, K.}, \bibinfo{author}{Puliafito, C.A.},
  \bibinfo{author}{Fujimoto, J.G.}, \bibinfo{year}{1991}.
\newblock \bibinfo{title}{{Optical Coherence Tomography}}.
\newblock \bibinfo{journal}{Science} \bibinfo{volume}{254},
  \bibinfo{pages}{1178--1181}.
\newblock \DOIprefix\doi{10.1126/science.1957169}.
\bibitem[{Krieger and Dougherty(1959)}]{Krieger:1959}
\bibinfo{author}{Krieger, I.M.}, \bibinfo{author}{Dougherty, T.J.},
  \bibinfo{year}{1959}.
\newblock \bibinfo{title}{{A Mechanism for Non-Newtonian Flow in Suspensions of
  Rigid Spheres}}.
\newblock \bibinfo{journal}{Transactions of The Society of Rheology}
  \bibinfo{volume}{3}, \bibinfo{pages}{137--152}.
\newblock \DOIprefix\doi{10.1122/1.548848}.
\bibitem[{Kundu et~al.(2025)Kundu, Usabiaga and Ellero}]{Kundu:2025}
\bibinfo{author}{Kundu, D.}, \bibinfo{author}{Usabiaga, F.B.},
  \bibinfo{author}{Ellero, M.}, \bibinfo{year}{2025}.
\newblock \bibinfo{title}{{Settling dynamics of a non-Brownian suspension of
  spherical and cubic particles in Stokes flow}}.
\newblock \bibinfo{journal}{Journal of Fluid Mechanics} \bibinfo{volume}{1010},
  \bibinfo{pages}{A63}.
\newblock \DOIprefix\doi{10.1017/jfm.2025.363}.
\bibitem[{Kynch(1952)}]{Kynch:1952}
\bibinfo{author}{Kynch, G.J.}, \bibinfo{year}{1952}.
\newblock \bibinfo{title}{{A theory of sedimentation}}.
\newblock \bibinfo{journal}{Trans. Faraday Soc.} \bibinfo{volume}{48},
  \bibinfo{pages}{166--176}.
\newblock \DOIprefix\doi{10.1039/tf9524800166}.
\bibitem[{Mueller et~al.(2009)Mueller, Llewellin and Mader}]{Mueller:2009}
\bibinfo{author}{Mueller, S.}, \bibinfo{author}{Llewellin, E.W.},
  \bibinfo{author}{Mader, H.M.}, \bibinfo{year}{2009}.
\newblock \bibinfo{title}{The rheology of suspensions of solid particles}.
\newblock \bibinfo{journal}{Proceedings of the Royal Society A: Mathematical,
  Physical and Engineering Sciences} \bibinfo{volume}{466},
  \bibinfo{pages}{1201--1228}.
\newblock \DOIprefix\doi{10.1098/rspa.2009.0445}.
\bibitem[{Mujat et~al.(2013)Mujat, Ferguson, Iftimia, Hammer, Nedyalkov, Wosnik
  and Legner}]{Mujat:2013}
\bibinfo{author}{Mujat, M.}, \bibinfo{author}{Ferguson, R.D.},
  \bibinfo{author}{Iftimia, N.}, \bibinfo{author}{Hammer, D.X.},
  \bibinfo{author}{Nedyalkov, I.}, \bibinfo{author}{Wosnik, M.},
  \bibinfo{author}{Legner, H.}, \bibinfo{year}{2013}.
\newblock \bibinfo{title}{{Optical coherence tomography-based micro-particle
  image velocimetry}}.
\newblock \bibinfo{journal}{Optics Letters} \bibinfo{volume}{38},
  \bibinfo{pages}{4558--4561}.
\newblock \DOIprefix\doi{10.1364/ol.38.004558}.
\bibitem[{Nicolai et~al.(1995)Nicolai, Herzhaft, Hinch, Oger and
  Guazzelli}]{Nicolai:1995}
\bibinfo{author}{Nicolai, H.}, \bibinfo{author}{Herzhaft, B.},
  \bibinfo{author}{Hinch, E.J.}, \bibinfo{author}{Oger, L.},
  \bibinfo{author}{Guazzelli, E.}, \bibinfo{year}{1995}.
\newblock \bibinfo{title}{{Particle velocity fluctuations and hydrodynamic
  self-diffusion of sedimenting non-Brownian spheres}}.
\newblock \bibinfo{journal}{Physics of Fluids} \bibinfo{volume}{7},
  \bibinfo{pages}{12--23}.
\newblock \DOIprefix\doi{10.1063/1.868733}.
\bibitem[{Pabst et~al.(2006)Pabst, Gregorová and Berthold}]{Pabst:2006}
\bibinfo{author}{Pabst, W.}, \bibinfo{author}{Gregorová, E.},
  \bibinfo{author}{Berthold, C.}, \bibinfo{year}{2006}.
\newblock \bibinfo{title}{{Particle shape and suspension rheology of
  short-fiber systems}}.
\newblock \bibinfo{journal}{Journal of the European Ceramic Society}
  \bibinfo{volume}{26}, \bibinfo{pages}{149--160}.
\newblock \DOIprefix\doi{10.1016/j.jeurceramsoc.2004.10.016}.
\bibitem[{Richardson and Zaki(1997)}]{Richardson:1997}
\bibinfo{author}{Richardson, J.}, \bibinfo{author}{Zaki, W.},
  \bibinfo{year}{1997}.
\newblock \bibinfo{title}{{Sedimentation and fluidisation: Part I}}.
\newblock \bibinfo{journal}{Chemical Engineering Research and Design}
  \bibinfo{volume}{75}, \bibinfo{pages}{S82--S100}.
\newblock \DOIprefix\doi{10.1016/s0263-8762(97)80006-8}.
\bibitem[{Ruhlandt and Salditt(2019)}]{Ruhlandt:2019}
\bibinfo{author}{Ruhlandt, A.}, \bibinfo{author}{Salditt, T.},
  \bibinfo{year}{2019}.
\newblock \bibinfo{title}{{Time-resolved x-ray phase-contrast tomography of
  sedimenting micro-spheres}}.
\newblock \bibinfo{journal}{New Journal of Physics} \bibinfo{volume}{21},
  \bibinfo{pages}{043017}.
\newblock \DOIprefix\doi{10.1088/1367-2630/ab13c8}.
\bibitem[{Segr\`{e} et~al.(1997)Segr\`{e}, Herbolzheimer and
  Chaikin}]{Segre:1997}
\bibinfo{author}{Segr\`{e}, P.N.}, \bibinfo{author}{Herbolzheimer, E.},
  \bibinfo{author}{Chaikin, P.M.}, \bibinfo{year}{1997}.
\newblock \bibinfo{title}{{Long-Range Correlations in Sedimentation}}.
\newblock \bibinfo{journal}{Physical Review Letters} \bibinfo{volume}{79},
  \bibinfo{pages}{2574--2577}.
\newblock \DOIprefix\doi{10.1103/physrevlett.79.2574}.
\bibitem[{Stickel and Powell(2005)}]{Stickel:2005}
\bibinfo{author}{Stickel, J.J.}, \bibinfo{author}{Powell, R.L.},
  \bibinfo{year}{2005}.
\newblock \bibinfo{title}{{Fluid Mechanics and Rheology of Dense Suspensions}}.
\newblock \bibinfo{journal}{Annual Review of Fluid Mechanics}
  \bibinfo{volume}{37}, \bibinfo{pages}{129--149}.
\newblock \DOIprefix\doi{10.1146/annurev.fluid.36.050802.122132}.
\bibitem[{Waitukaitis and Jaeger(2012)}]{Waitukaitis:2012}
\bibinfo{author}{Waitukaitis, S.R.}, \bibinfo{author}{Jaeger, H.M.},
  \bibinfo{year}{2012}.
\newblock \bibinfo{title}{{Impact-activated solidification of dense suspensions
  via dynamic jamming fronts}}.
\newblock \bibinfo{journal}{Nature} \bibinfo{volume}{487},
  \bibinfo{pages}{205--209}.
\newblock \DOIprefix\doi{10.1038/nature11187}.
\bibitem[{Zhou et~al.(2016)Zhou, Huang, Gamm, Bhandari, Khokha and
  Choma}]{Zhou:2016}
\bibinfo{author}{Zhou, K.C.}, \bibinfo{author}{Huang, B.K.},
  \bibinfo{author}{Gamm, U.A.}, \bibinfo{author}{Bhandari, V.},
  \bibinfo{author}{Khokha, M.K.}, \bibinfo{author}{Choma, M.A.},
  \bibinfo{year}{2016}.
\newblock \bibinfo{title}{{Particle streak velocimetry-optical coherence
  tomography: a novel method for multidimensional imaging of microscale fluid
  flows}}.
\newblock \bibinfo{journal}{Biomedical Optics Express} \bibinfo{volume}{7},
  \bibinfo{pages}{1590--1603}.
\newblock \DOIprefix\doi{10.1364/boe.7.001590}.

\end{thebibliography}

\end{document}